\documentclass[aip,reprint,amsmath,amssymb,floatfix,citeautoscript]{revtex4-2}
\usepackage[utf8]{inputenc}
\usepackage{graphicx}
\usepackage{color}
\usepackage{bm}
\usepackage{braket}
\usepackage[colorlinks,linkcolor=blue,citecolor=blue,urlcolor=blue]{hyperref}

\begin{document}

\title{Dielectric Loss due to Charged-Defect Acoustic Phonon Emission}

\author{Mark E. Turiansky}
\email{mturiansky@physics.ucsb.edu}
\affiliation{Materials Department, University of California, Santa Barbara, CA 93106-5050, U.S.A.}

\author{Chris G. \surname{Van de Walle}}
\email{vandewalle@mrl.ucsb.edu}
\affiliation{Materials Department, University of California, Santa Barbara, CA 93106-5050, U.S.A.}

\date{\today}

\begin{abstract}
    The coherence times of state-of-the-art superconducting qubits are limited by bulk dielectric loss, yet the microscopic mechanism leading to this loss is unclear.
    Here we propose that the experimentally observed loss can be attributed to the presence of charged defects that enable the absorption of electromagnetic radiation by the emission of acoustic phonons.
    Our explicit derivation of the absorption coefficient for this mechanism allows us to derive a loss tangent of $7.2 \times 10^{-9}$
    for Al$_2$O$_3$, in good agreement with recent high-precision measurements [A. P. Read {\it et al.}, Phys. Rev. Appl. {\bf 19}, 034064 (2023)].
    We also find that for temperatures well below $\sim$0.2 K, the loss should be independent of temperature, also in agreement with observations.
    Our investigations show that the loss per defect depends mainly on properties of the host material, and a high-throughput search suggests that diamond, cubic BN, AlN, and SiC are optimal in this respect.
\end{abstract}

\maketitle

\section{Introduction}

Superconducting qubits are currently the main hardware for realizing quantum computing~\cite{arute_quantum_2019}.
A key figure of merit is the coherence time, which describes how long the qubit can retain quantum information.
Extending coherence times is imperative to enable scaling up beyond hundreds of qubits and realizing universal quantum computing.
Progress has been made by feats of engineering, such as exploring different device architectures, implementing control protocols, and learning to work with noise~\cite{kjaergaard_superconducting_2020,krasnok_advancements_2023}.
The coherence times are now reaching the fundamental limits set by the materials used to construct the devices~\cite{oconnell_microwave_2008,krasnok_advancements_2023}.
Improvements in materials are thus essential to extend coherence times.

Identifying the microscopic loss mechanisms that limit coherence times is challenging~\cite{de_leon_materials_2021,mcrae_materials_2020}.
Surfaces and interfaces have been implicated as a source of loss~\cite{wang_surface_2015,woods_determining_2019}
Judicious use of cleaning techniques~\cite{place_new_2021} and device architectures that minimize surface participation~\cite{crowley_disentangling_2023}
have led to marked improvements, to the point where bulk dielectric loss is now limiting state-of-the-art devices based on Al$_2$O$_3$~\cite{read_precision_2023},
which is the most commonly used substrate for current qubits~\cite{de_leon_materials_2021}.

The microscopic mechanism that leads to this bulk dielectric loss is, however, still unclear.
Two-level systems (TLSs) have received the most attention, both experimentally and theoretically~\cite{muller_towards_2019}.
However, TLSs give rise to resonances, while the spectrum of the dominant loss is broad and featureless.
Intrinsic bulk dielectric loss is expected to be exceedingly low in a perfect crystal~\cite{gurevich_intrinsic_1991}.
However, even the highest-quality crystals contain some concentration of defects (we use the term ``defect'' to refer to both intrinsic point defects and extrinsic impurities).
Point defects are of course known to be able to absorb light or infrared radiation~\cite{stoneham_theory_1975},
but the ability for defects to absorb microwave radiation has been overlooked.

Here we explore a mechanism that can explain absorption of microwave radiation by charged defects via the emission of acoustic phonons~\cite{garin_one-phonon_1990,gurevich_intrinsic_1991}.
We develop a formalism that allows for quantitative evaluation of absorption coefficients and loss tangents, and show that this mechanism
can explain the observed dielectric loss in high-quality Al$_2$O$_3$ crystals.
Furthermore, the temperature dependence (or lack thereof) agrees with experimental observations.

Given the ability of our formalism to describe experimentally observed loss, we implement a high-throughput search for alternative materials with low loss.
The results show that diamond, cubic BN, AlN, and SiC are promising host for limiting these dielectric losses and thus
further improving the coherence times of superconducting qubits.

\section{Theory}
\label{sec:theory}

Dielectric loss must be due to an absorption mechanism in which energy is dissipated, e.g., in the form of phonons.
Superconducting qubits are typically operated between 1-10~GHz~\cite{kjaergaard_superconducting_2020}.
Optical phonon modes, with typical frequencies on the order of THz, are too far off resonance to contribute to absorption at these frequencies.
Acoustic phonon modes, on the other hand, form a continuum with frequencies ranging from zero to THz.
They are therefore prime candidates to lead to absorption in the microwave regime, provided there is a mechanism to couple to radiation.

Consider first a perfect crystal in the absence of defects.
Photons carry (approximately) no momentum, meaning that only zone-center phonon modes can be excited.
Therefore, in a perfect crystal with translational symmetry, acoustic phonon modes cannot lead to absorption to first order~\cite{lax_infrared_1955}.

The introduction of defects into the crystal has two important consequences.
First, assuming the defects are randomly distributed, the translational symmetry of the crystal will be broken.
This relaxes the zone-center requirement, as all phonon modes can be thought of as being folded back to the zone center in the imperfect crystal; a phonon mode from anywhere in the Brillouin zone can then in principle contribute to absorption if it is on resonance.
Second, if the defect is charged, the displacement of charge produces a dipole that can couple to electromagnetic radiation.

We want to calculate the absorption coefficient $a (\omega)$ at photon frequency $\omega$ in the presence of a charged defect.
The absorption coefficient is related to the attenuation of the electromagnetic radiation in the material by~\cite{stoneham_theory_1975}
\begin{equation}
    \label{eq:I}
    I(z) = I(0) \, e^{-a(\omega) z} \;,
\end{equation}
where $I$ is the intensity at coordinate $z$ along the direction of propagation.
$a(\omega)$ can be related to the absorption cross section per defect $\sigma (\omega)$ by
\begin{equation}
    \label{eq:ktosigma}
    a(\omega) = N_{\rm def} \, \sigma(\omega) \;.
\end{equation}
where $N_{\rm def}$ is the density of defects assumed to be distributed homogeneously throughout the material.

The absorption cross section $\sigma(\omega)$ can be obtained by application of Fermi's golden rule,
\begin{multline}
    \label{eq:sigma}
    \sigma (\omega) = Z_{\rm eff}^2 {\left(\frac{\mathcal{E}_{\rm eff}}{\mathcal{E}_0} \right)}^2 \frac{4 \pi^2 \alpha}{n_r} \times \\
    \sum_{I,F} p_I {\lvert \braket{\Phi_F \lvert {\bm \xi} \cdot \hat{\bm u}_d \rvert \Phi_I} \rvert}^2 \Omega_{IF} \delta (\omega - \Omega_{IF}) \;,
\end{multline}
where $\alpha$ is the fine-structure constant, $n_r$ is the index of refraction, $p_I$ is the thermal occupation factor, and $\hbar\Omega_{IF}$ is the energy difference between the initial and final states.
The prefactor $(\mathcal{E}_{\rm eff} / \mathcal{E}_0)^2$ represents local-field effects, where $\mathcal{E}_{\rm eff}$ is the effective electric field at the defect and $\mathcal{E}_0$ the bulk value.
This factor captures the fact that the electric field at the defect is different from that in bulk due to scattering of light at the defect~\cite{stoneham_theory_1975}.
Various models~\cite{smith_v_1972,stoneham_theory_1975} have been proposed to model the effect.
In the Lorentz-Lorenz model~\cite{lorentz_theory_1916}, $\mathcal{E}_{\rm eff} / \mathcal{E}_0 = (\varepsilon + 2)/3$ where $\varepsilon$ is the dielectric permittivity.
In the Onsager model~\cite{onsager_electric_1936}, $\mathcal{E}_{\rm eff} / \mathcal{E}_0 = 3 \varepsilon /(2 \varepsilon +1)$.
The Lorentz-Lorenz model is generally considered to be an overestimate of local-field effects;
in the present work we therefore use the Onsager model, recognizing that it probably represents an underestimate of the local-field enhancement.

${\bm \xi}$ is the polarization vector of the incoming radiation.
$Z_{\rm eff}$ is the Born effective charge; for coupling to an acoustic phonon mode this corresponds to the charge on the defect~\cite{cockayne_influence_2007} in the usual treatment of charge states of defects~\cite{freysoldt_first-principles_2014}.
$\hat{\bm u}_d$ is the displacement operator for the defect site, whose evaluation will be discussed shortly.
$\ket{\Phi_J}$ are the many-body phonon wavefunctions, which are given by direct products of the usual harmonic oscillator wavefunctions $\ket{\phi_{k,n_k}}$:
\begin{equation}
    \label{eq:wf}
    \ket{\Phi_J} = \prod_{k=1}^{3N_a} \ket{\phi_{k,n_k}} \;,
\end{equation}
where $N_a$ is the number of atoms in the crystal and $n_k$ is the occupation number of the $k$th phonon mode.
Superconducting qubits typically operate at mK temperatures;
we will assume that all phonon modes are in the $n_k = 0$ state in the initial state.
We can then remove the summation over $I$ in Eq.~(\ref{eq:sigma}) and set $p_0 = 1$.

\begin{figure}[htb!]
    \centering
    \includegraphics[width=\columnwidth,height=0.5\textheight,keepaspectratio]{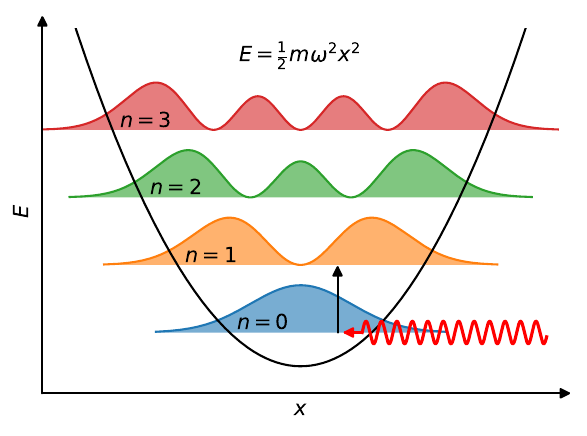}
    \caption{\label{fig:schematic}
        The potential energy surface of a phonon mode in the harmonic approximation.
        The probability densities of the vibronic wavefunctions are shaded in color.
        At low $T$ only the $n = 0$ harmonic oscillator level is occupied and microwave radiation (wavy red arrow) drives a transition to the $n = 1$ level.
    }
\end{figure}

We are concerned with a purely vibronic transition (schematically depicted in Fig.~\ref{fig:schematic}), but it can be useful to draw an analogy with the case of an electronic transition.
For an electronic transition, the vibronic wavefunctions $\Phi_J$ is replaced with the electronic wavefunctions $\Psi_J$, and
the defect displacement operator $\hat{\bm u}_d$ is replaced with the position operator $\hat{\bm r}$.
The dipole for an electronic transition would then be $e \lvert \braket{\Psi_F \lvert {\bm \xi} \cdot \hat{\bm r} \rvert \Psi_I} \rvert$, compared to $e Z_{\rm eff} \lvert \braket{\Phi_F \lvert {\bm \xi} \cdot \hat{\bm u}_d \rvert \Phi_I} \rvert$ for a vibronic transition.

The displacement $\hat{\bm u}_d$ is not a normal mode of the system, but can be expanded in the basis of normal modes,
\begin{equation}
    \label{eq:ud}
    {\bm \xi} \cdot \hat{\bm u}_d = \sum_{j=1}^{3N_a} \chi_j \, \sqrt{\frac{\hbar}{2 M_j \omega_j}} \left( \hat{a}_j^\dagger + \hat{a}_j \right) \;,
\end{equation}
where $a_j$ ($a_j^\dagger$) is the annihilation (creation) operator for the phonon mode with frequency $\omega_j$ and modal mass $M_j$,
and $\chi_j$ are the expansion coefficients.
Since the initial state has $n_k = 0$, the annihilation operator in Eq.~(\ref{eq:ud}) does not contribute,
and only final states that differ from the initial state by the occupation of a single phonon mode can contribute.
Equation~(\ref{eq:sigma}) then reduces to
\begin{equation}
    \label{eq:sigma0}
    \sigma (\omega) = Z_{\rm eff}^2 {\left(\frac{\mathcal{E}_{\rm eff}}{\mathcal{E}_0} \right)}^2 \frac{4 \pi^2 \alpha}{n_r} \sum_{j=1}^{3N_a} \frac{\hbar}{2 M_j} {\lvert \chi_j \rvert}^2 \delta (\omega - \omega_j) \;.
\end{equation}
We are left with the task of determining the expansion coefficients $\chi_j$.

In principle, the expansion coefficients could be obtained directly from a supercell calculation using, e.g., density functional perturbation theory~\cite{gonze_dynamical_1997}.
While this is practical for optical phonons, this approach is infeasible for acoustic phonons due to their extended nature.
For example, since the speed of sound in silicon is $\sim$8400 m/s, an acoustic phonon with a 4~GHz frequency has a wavelength of 2.1~$\mu$m.
It is impossible to perform a first-principles calculation for a supercell of this size.
A model calculation is called for, and we follow the approach of Dawber and Elliott~\cite{dawber_vibrations_1962,dawber_theory_1963}
who studied a ``mass defect'', where only the mass of the defect atom is changed and force constants retain their bulk values.
For the case of a monatomic cubic crystal they found
\begin{equation}
    \label{eq:dawber_elliot}
    {\lvert \chi_j \rvert}^2 = \frac{1}{N_a} {\left\{ \pi^2 \epsilon^2 z^2 \nu^2 (z) + {\left[ 1 + \epsilon z \; \mathcal{P} \!\! \int \frac{d\mu \, \nu(\mu)}{\mu - z} \right]}^2 \right\}}^{-1} \;,
\end{equation}
where $\mu = \omega^2$, $z = \omega_j^2$, and $\mathcal{P}$ indicates that we take the Cauchy principal value of the integral.
$\epsilon = (M - M^\prime) / M$ is a measure of the defect mass $M^\prime$ with respect to the atomic mass $M$ of the lattice species.
$\nu (\mu)$ is the density of phonon states per squared unit frequency with normalization
\begin{equation}
    \label{eq:nu_norm}
    \int d\mu \, \nu(\mu) = 1 \;.
\end{equation}
[I.e., $\nu (\omega^2) = \rho(\omega) / 2\omega$, where $\rho(\omega)$ is the usual density of phonon states per unit frequency.]

In the low frequency limit ($z = \omega_j^2 \rightarrow 0$) relevant for acoustic phonons, Eq.~(\ref{eq:dawber_elliot}) reduces to
\begin{equation}
    \label{eq:dawber_elliot_acoustic}
    {\lvert \chi_j \rvert}^2 = \frac{1}{3 N_a} \;.
\end{equation}
One could have obtained this result by noting that the displacement of the defect occurs on a single site, while the acoustic phonons are delocalized over all $N_a$ atoms of the crystal.
Indeed acoustic phonons are insensitive to the microscopic details and are effectively a bulk material property.
We thus expect this to be a robust result, independent of the model or calculation method.
In particular, the following results are not restricted to the case of a monatomic cubic crystal.

We can now evaluate the absorption cross section, Eq.~(\ref{eq:sigma0}).
Since the phonons make up a continuum we convert the summation over final states to an integral,
\begin{equation}
    \sum_{j=1}^{3N_a} \rightarrow 3N_a \int d\omega_j \, \rho (\omega_j) \;,
\end{equation}
where $\rho(\omega_j)$ is the phonon density of states normalized to unity.
With Eq.~(\ref{eq:dawber_elliot_acoustic}) we can then write the cross section as
\begin{equation}
    \label{eq:sigma1}
    \sigma (\omega) = Z_{\rm eff}^2 {\left(\frac{\mathcal{E}_{\rm eff}}{\mathcal{E}_0} \right)}^2 \frac{4 \pi^2 \alpha}{n_r} \frac{\hbar}{2 M} \rho(\omega) \;,
\end{equation}
where we have replaced the modal mass $M_j$ by the average atomic mass of the host lattice $M$, which is appropriate for acoustic phonons.

Finally, we use the fact that the phonon density of states for acoustic phonons is well described by the Debye model:
\begin{equation}
    \label{eq:debye}
    \rho(\omega) = \frac{3 \omega^2}{\omega_m^3} \;.
\end{equation}
$\omega_m$ is the Debye frequency, which is the maximum possible phonon frequency in the model [i.e., $\rho(\omega > \omega_m) = 0$], and is given by
\begin{equation}
    \label{eq:w_max}
    \omega_m = 2 \pi {\left( \frac{3 N_s}{4 \pi} \right)}^{1/3} v_s \;,
\end{equation}
where $N_s$ is the site density and $v_s$ is the sound velocity.
The sound velocity can be obtained from the elastic constants of the material,
\begin{align}
    \label{eq:sound}
    v_s^{(t)} = \sqrt{\frac{G}{\varrho}} \;, 
    v_s^{(l)} = \sqrt{\frac{K + 4G/3}{\varrho}} \;,
\end{align}
where $K$ is the bulk modulus, $G$ is the shear modulus, and $\varrho$ is the mass density.
The superscript indicates whether the speed corresponds to the transverse $(t)$ or longitudinal $(l)$ mode.
Our goal is to reproduce the low-frequency phonon density of states, and
we find from empirically fitting the full phonon density of states that the transverse component is close to the fitted value.
Thus we will use the transverse speed of sound in subsequent calculations.

Plugging Eq.~(\ref{eq:sigma1}) into Eq.~(\ref{eq:ktosigma}) with the Debye model for the density of states [Eq.~(\ref{eq:debye})] leads to an absorption coefficient
\begin{equation}
    \label{eq:abs_simple}
    a(\omega) = \underbrace{N_{\rm def} Z_{\rm eff}^2 \vphantom{\frac{\omega^2}{\omega_m^3}}}_\text{defect} \underbrace{{\left(\frac{\mathcal{E}_{\rm eff}}{\mathcal{E}_0} \right)}^2 \frac{6 \pi^2 \alpha}{n_r} \frac{\hbar}{M} \frac{\omega^2}{\omega_m^3}}_\text{host} \;,
\end{equation}
where the brackets highlight the fact that there are two separate contributions:
(i) a defect-specific contribution described by the density of defects $N_{\rm def}$ and effective charge $Z_{\rm eff}$,
and (ii) a host-specific contribution described by the index of refraction $n_r$, average atomic mass $M$, and Debye frequency $\omega_m$.
In this light, $\sigma(\omega) / Z_{\rm eff}^2$ can be viewed as the ``intrinsic'' cross section that only depends on the host material parameters.
We can define a characteristic parameter $\mathcal{A}_{\rm c}$, given by
\begin{equation}
    \label{eq:ac}
    \mathcal{A}_c = \frac{\mathcal{E}_{\rm eff}}{\mathcal{E}_0} \sqrt{\frac{6 \pi^2 \alpha}{n_r} \frac{\hbar}{M} \frac{1}{\omega_m^3}} \;,
\end{equation}
which depends only on the host material and has units of an absement (m$\cdot$s).
The final absorption coefficient then takes the convenient form
\begin{equation}
    \label{eq:abs_fin}
    a(\omega) = N_{\rm def} Z_{\rm eff}^2 \mathcal{A}_c^2 \omega^2 \;.
\end{equation}

Experimental measurements typically report a loss tangent $\tan(\delta)$, which is the ratio of the imaginary part of the dielectric permittivity $\epsilon(\omega)$ (which is related to the absorption coefficient) to the real part:
\begin{equation}
    \label{eq:loss_tan}
    \tan(\delta) = \frac{{\rm Im} \, \epsilon(\omega)}{{\rm Re} \, \epsilon(\omega)} = \frac{c}{n_r \omega} a(\omega) \;,
\end{equation}
where we have used ${\rm Re} \, \epsilon (\omega) \approx n_r^2$.
When $\delta$ is small, $\tan(\delta) \approx \delta$.
Utilizing Eqs.~(\ref{eq:w_max}) and (\ref{eq:abs_simple}), we obtain
\begin{equation}
    \label{eq:loss_tan_garin}
    \tan(\delta) = \frac{1}{4\pi\epsilon_0n_r^2} {\left(\frac{\mathcal{E}_{\rm eff}}{\mathcal{E}_0} \right)}^2 \frac{N_{\rm def} Z_{\rm eff}^2 e^2}{\varrho v_s^3} \omega \;.
\end{equation}
This equation is very similar to the result from Garin~\cite{garin_one-phonon_1990}, despite starting from different equations.
Garin started from an elastic continuum model, focusing only on the acoustic phonon modes.
Garin included the effect of spatial correlations between the defects, obtaining a dependence that can be viewed as a correction factor to our result.
Here we assume that defect positions are random and uncorrelated, which is consistent with dilute defect concentrations and means no correction is necessary.
Furthermore, Garin neglected local-field effects.

For purposes of distinguishing this loss mechanism from others, it is useful to look at the scaling as a function of various parameters.
Eq.~(\ref{eq:loss_tan_garin}) clearly shows that the loss is proportional to the frequency $\omega$.
While we did not explicitly include the temperature dependence in the formulas, we can make some comments.
When the temperature is low relative to the phonon frequency ($k T \ll \hbar\omega$), we can assume that the phonons are in their ground state and our derivation holds.
Therefore, as long as $k T$ stays well below $\hbar\omega$, the absorption will be independent of temperature.
For $\omega / (2\pi) = 4$~GHz, this corresponds to a temperature of 192~mK.

\section{Results}

Now that we have explicit equations we can turn to quantitatively evaluating the loss.
In principle, the values that are necessary to evaluate Eq.~(\ref{eq:ac}) could be obtained from experiments.
However, we would like to assess a wide range of materials, and a curated experimental database that contains all the necessary parameters does not exist to our knowledge.
Towards this end, we will utilize high-throughput first-principles calculations based on density functional theory.
We will utilize the database from the Materials Project~\cite{jain_commentary_2013}, which at the writing of this manuscript contains 154,718 materials.

The characteristic parameter $\mathcal{A}_c$ depends on the average mass $M$, which is trivially obtained from the host crystal structure, and the Debye frequency $\omega_m$, which is obtained from the elastic constants of the material [Eqs.~(\ref{eq:w_max}) and (\ref{eq:sound})].
The index of refraction can be approximately obtained from the dielectric response of the material:
\begin{equation}
    \label{eq:nr}
    n_r = \sqrt{{\rm Tr} [\overleftrightarrow{\varepsilon}] / 3} \;,
\end{equation}
where $\overleftrightarrow{\varepsilon}$ is the dielectric tensor of the material.
In principle, the index of refraction should be evaluated at the frequency of interest, but the frequency-dependent value is not easily accessible.
Using the low-frequency value of the dielectric tensor is expected to be a good approximation; indeed, experimental values for dielectric permittivities of Al$_2$O$_3$ at 8 GHz (Ref.~\onlinecite{krupka_use_1999}) are in very good agreement with values obtained at audio frequencies (Ref.~\onlinecite{shelby_low_1980}).
All these parameters are amenable to high-throughput simulation, since they can be obtained with reasonable accuracy using a semi-local functional.
Elastic properties are known to be well described by semi-local functionals.
While the band gap is underestimated by semi-local functionals, the deviations for dielectric properties from experiment are within a reasonable uncertainty.

We first examine the case of Al$_2$O$_3$, for which experimental results are available~\cite{read_precision_2023}.
Read \textit{et al.}~\cite{read_precision_2023} performed highly sensitive measurements of the loss tangent in high-grade Al$_2$O$_3$ grown with the heat-exchanger method (HEMEX), measuring a loss tangent of $19 \times 10^{-9}$ at 4.5~GHz.
The dominant impurity in HEMEX Al$_2$O$_3$ has been identified to be Si, which is present in concentrations on the order of $10^{18}$~cm$^{-3}$~\cite{alexandrovski_absorption_2001,route_heat-treatment_2004,prot_self-diffusion_1996}.
Silicon prefers to incorporate on the Al site in Al$_2$O$_3$ and acts as a donor in a +1 charge state~\cite{mu_role_2022};
charge neutrality then requires acceptors to be present in similar concentrations.
Alumininum vacancies are the most likely compensating species, acting as deep acceptors predominantly in the $-3$ charge state~\cite{choi_native_2013}.
Thus we will use $Z_{\rm eff} = +1$ and $N_{\rm def} = 10^{18}$~cm$^{-3}$ for Si, $Z_{\rm eff} = -3$ and $N_{\rm def} = 1/3 \times 10^{18}$~cm$^{-3}$ for $V_{\rm Al}$, and add their contributions.
With the values for the Debye frequency and index of refraction calculated from first principles for Al$_2$O$_3$, we find a loss tangent of $\tan(\delta) = 7.2 \times 10^{-9}$.
Experimentally a value of $19 \times 10^{-9}$ (Ref.~\onlinecite{read_precision_2023}) was found.

There are several reasons why our calculated value is expected to be an underestimate of the experimental value.
First, contributions from other defects are likely to be present.
Indeed, Refs.~\onlinecite{alexandrovski_absorption_2001} and \onlinecite{route_heat-treatment_2004} reported a number of other impurities to be present in concentrations on the order of $10^{17}$~cm$^{-3}$, and each charged impurity will be accompanied by compensating point defects such as Al vacancies.
Second, as discussed in Sec.~\ref{sec:theory}, our treatment of local-field effects in the Onsager model likely represents an underestimate, possibly by as much as an order of magnitude.
Within the Onsager model, the enhancement factor $(\mathcal{E}_{\rm eff} / \mathcal{E}_0 )^2$ for Al$_2$O$_3$ is equal to 2.04, while in
the Lorentz-Lorenz model the factor would be 15.3.

We also note that Read \textit{et al}. found the loss tangent to be independent of temperature below 200~mK, as predicted by our model.
Overall, the agreement found here validates our hypothesis that absorption by charged defects is the limiting factor in bulk loss in these samples.

The concentration of defects will depend on the growth method used.
The HEMEX sample examined in Ref.~\onlinecite{read_precision_2023} was actually the highest grade Al$_2$O$_3$ available.
Read \textit{et al}. also measured Al$_2$O$_3$ produced by edge-defined film-fed growth (EFG) and found the loss tangent to be a factor of three higher, $63 \times 10^{-9}$~\cite{read_precision_2023}.
While we were not able to find measurements of impurity concentrations in EFG samples, it is plausible that defect concentrations are a factor of 3 higher in EFG Al$_2$O$_3$ compared to HEMEX.
Later studies found that annealing EFG Al$_2$O$_3$ at 1473~K resulted in loss tangents similar to that of HEMEX material~\cite{ganjam_surpassing_2023}.
Annealing at such temperatures may allow defects to migrate and recombine or form complexes, thus lowering the loss.

If absorption by charged defects is the limiting factor in bulk loss, then controlling this loss is of paramount importance for improving coherence times of superconducting qubits.
Since the loss is proportional to the defect concentration, controlling the incorporation of defects using high-quality and high-purity crystal growth techniques is essential.
However, our derivation of the defect-induced absorption [Eq.~(\ref{eq:abs_fin})] also makes clear that the loss depends on properties of the host material.
It is therefore fruitful to identify materials that may have lower loss than Al$_2$O$_3$, based on a high-throughput search.
Of the materials in the Materials Project database, we focus on those that have a finite band gap when using the semi-local functional of Perdew, Burke, and Ernzerhof (PBE)~\cite{perdew_generalized_1996}, as well as having their elastic~\cite{de_jong_charting_2015} and dielectric properties~\cite{petousis_high-throughput_2017} calculated.
Furthermore, we exclude magnetic materials from the search to avoid the presence of spins that may contribute magnetic noise.

\begin{figure*}[htb!]
    \centering
    \includegraphics[width=\textwidth,height=3.5in,keepaspectratio]{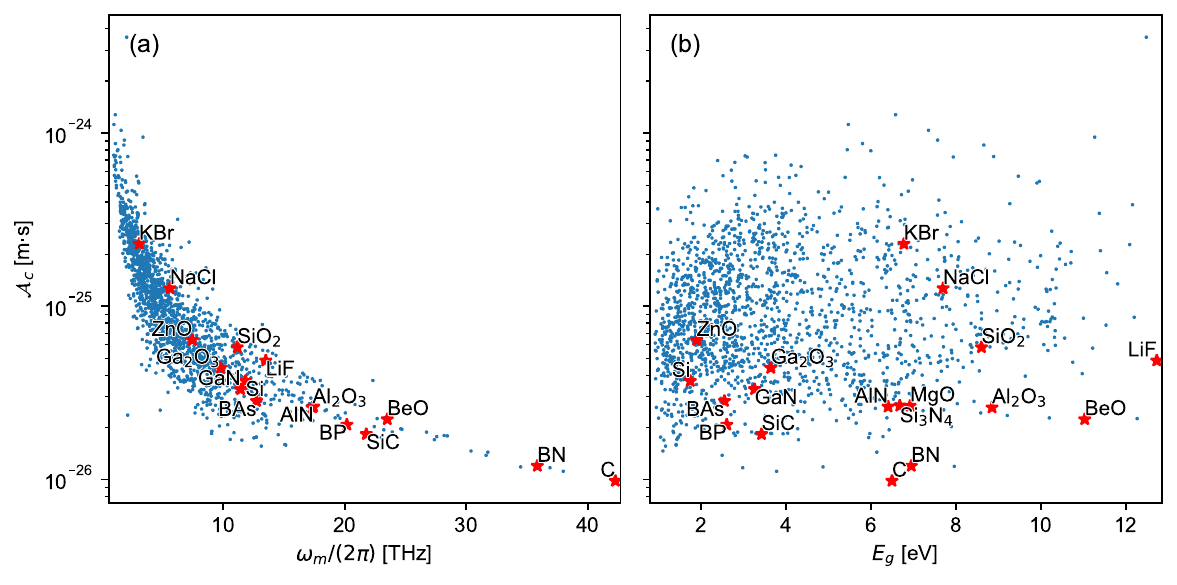}
    \caption{\label{fig:ac}
        High-throughput calculations of the characteristic parameter $\mathcal{A}_c$ [Eq.~(\ref{eq:ac})] as a function of (a) the Debye frequency $\omega_m$ [Eq.~(\ref{eq:w_max})] and (b) the band gap $E_g$ [Eq.~(\ref{eq:eg})].
    }
\end{figure*}

The characteristic parameters $\mathcal{A}_c$ of the remaining 1,821 materials are shown in Fig.~\ref{fig:ac}.
Values for selected materials are shown in Table~\ref{tab:results}, along with the loss tangent $\tan(\delta)$ calculated assuming $Z_{\rm eff} = 1$ and $N_{\rm def} = 10^{18}$~cm$^{-3}$.
As expected from Eq.~(\ref{eq:ac}), there is a strong dependence on the Debye frequency $\omega_m$ [Fig.~\ref{fig:ac}(a)].
Indeed this explains why diamond has the lowest $\mathcal{A}_c$ of any material in the database.
We also show the dependence of $\mathcal{A}_c$ on the band gap of the material [Fig.~\ref{fig:ac}(b)].
Since it is known that semi-local functionals underestimate the band gap~\cite{perdew_density_1985}, we utilize the empirical correction of Morales-Garc{\'i}a \textit{et al}.~\cite{morales-garcia_empirical_2017},
\begin{equation}
    \label{eq:eg}
    E_g = 1.355 \times E_g^{\rm PBE} + 0.916 \;,
\end{equation}
where $E_g^{\rm PBE}$ is the band gap computed at the PBE level.
There is no explicit dependence of $\mathcal{A}_c$ on the band gap in our equations;
however, the band gap can be an important selection criterion for evaluating material choices.
For example, silicon was chosen as a barrier material for the recently proposed FinMET design~\cite{goswami_towards_2022} because the smaller band gap would enable thicker barriers to be used.
We will further discuss the importance of the band gap in the following section.

\begin{table*}
    \caption{\label{tab:results}
        Debye frequency $\omega_m$, index of refraction $n_r$, characteristic parameter $\mathcal{A}_c$, and loss tangent $\tan(\delta)$ for selected materials from the high-throughput search.
        $\tan (\delta)$ is calculated assuming $\omega/(2\pi) = 4.5$~GHz, $N_{\rm def} = 10^{18}$~cm$^{-3}$, and $Z_{\rm eff} = 1$.
    }
    \begin{ruledtabular}
        \begin{tabular}{ccccccc}
Material & Space Group & Centrosymmetric? & $\omega_m/(2\pi)$ [THz] & $n_r$ & $\mathcal{A}_c$ [m$\cdot$s] & $\tan (\delta)$ \\
\hline
C & $Fd\bar{3}m$ & Y & 42.3 & 2.41 & $9.8 \times 10^{-27}$ & $3.4 \times 10^{-10}$ \\
BN & $F\bar{4}3m$ & N & 35.8 & 2.64 & $1.2 \times 10^{-26}$ & $4.6 \times 10^{-10}$ \\
SiC & $R3m$ & N & 21.8 & 3.29 & $1.8 \times 10^{-26}$ & $8.6 \times 10^{-10}$ \\
BP & $F\bar{4}3m$ & N & 20.2 & 3.05 & $2.1 \times 10^{-26}$ & $1.2 \times 10^{-9}$ \\
BeO & $P6_3mc$ & N & 23.5 & 2.71 & $2.2 \times 10^{-26}$ & $1.6 \times 10^{-9}$ \\
MgO & $Fm\bar{3}m$ & Y & 16.9 & 3.28 & $2.7 \times 10^{-26}$ & $1.8 \times 10^{-9}$ \\
Al$_{2}$O$_{3}$ & $R\bar{3}c$ & Y & 17.3 & 3.12 & $2.6 \times 10^{-26}$ & $1.8 \times 10^{-9}$ \\
AlN & $P6_3mc$ & N & 17.4 & 2.95 & $2.6 \times 10^{-26}$ & $2.0 \times 10^{-9}$ \\
Si$_{3}$N$_{4}$ & $P6_3/m$ & Y & 17.6 & 2.86 & $2.7 \times 10^{-26}$ & $2.1 \times 10^{-9}$ \\
BAs & $F\bar{4}3m$ & N & 12.7 & 3.15 & $2.8 \times 10^{-26}$ & $2.2 \times 10^{-9}$ \\
GaN & $P6_3mc$ & N & 11.4 & 3.30 & $3.3 \times 10^{-26}$ & $2.8 \times 10^{-9}$ \\
Si & $Fd\bar{3}m$ & Y & 11.8 & 3.61 & $3.7 \times 10^{-26}$ & $3.2 \times 10^{-9}$ \\
Ga$_{2}$O$_{3}$ & $C2/m$ & Y & 9.8 & 3.34 & $4.4 \times 10^{-26}$ & $4.9 \times 10^{-9}$ \\
LiF & $Fm\bar{3}m$ & Y & 13.5 & 2.95 & $4.9 \times 10^{-26}$ & $6.8 \times 10^{-9}$ \\
ZnO & $P6_3mc$ & N & 7.5 & 3.29 & $6.4 \times 10^{-26}$ & $1.1 \times 10^{-8}$ \\
SiO$_{2}$ & $P3_221$ & N & 11.2 & 2.18 & $5.8 \times 10^{-26}$ & $1.3 \times 10^{-8}$ \\
NaCl & $Fm\bar{3}m$ & Y & 5.6 & 2.56 & $1.3 \times 10^{-25}$ & $5.3 \times 10^{-8}$ \\
KBr & $Fm\bar{3}m$ & Y & 3.1 & 2.20 & $2.3 \times 10^{-25}$ & $2.0 \times 10^{-7}$ \\
        \end{tabular}
    \end{ruledtabular}
\end{table*}

\section{Discussion}

Silicon has also been used in constructing superconducting qubits~\cite{oconnell_microwave_2008,kim_enhanced_2021,goswami_towards_2022}.
Our predicted loss tangent for Si is slightly higher than for Al$_2$O$_3$, assuming equivalent defect concentrations.
However, given the importance of Si for semiconductor technology, defect control is very advanced and defect concentrations far lower than the $10^{18}$~cm$^{-3}$ value assumed in Table~\ref{tab:results} are readily achievable.
Thus one would expect the dielectric loss from charged defects in Si to be far less than that in Al$_2$O$_3$, and it could potentially be reduced further in specially prepared samples.

Surprisingly, measurements on devices based on Si measure loss tangents an order of magnitude greater than those for Al$_2$O$_3$~\cite{woods_determining_2019,melville_comparison_2020}.
Recent measurements of loss tangents in high-resistivity Si using high-Q superconducting resonators found a value two orders of magnitude greater than that of Al$_2$O$_3$~\cite{checchin_measurement_2022}.
Checchin {\it et al.}~\cite{checchin_measurement_2022} attribute the dominant loss to residual conductivity arising from variable-range hopping between defects.
This highlights the importance of the band gap as a design criterion [Fig.~\ref{fig:ac}(b)].
Narrow band-gap materials tend to host shallow defects, which have loosely bound hydrogenic wavefunctions that are easy to ionize.
Ultra-wide band-gap materials, on the other hand, tend to not have shallow defects and instead are dominated by deep defects, which have tightly bound wavefunctions and are harder to ionize.
(In other words, this is the difference between a semiconductor and an insulator.)
We can hypothesize that in the absence of this conduction mechanism, Si would have had lower loss due to better control over deep defects.

Piezoelectricity has also been invoked as a potential mechanism that may lead to loss~\cite{ioffe_decoherence_2004}.
In a similar way to the presence of charged defects, piezoelectricity enables the conversion of electromagnetic radiation into acoustic phonons.
To our knowledge, there is no rigorous confirmation of this proposal, except empirical evidence that switching from wurtzite to zincblende AlN improved coherence times in some devices~\cite{nakamura_superconducting_2011}.
Zincblende AlN is still piezoelectric but has zero coupling for certain orientations of the electric field.
Thus one an engineer around piezoelectricity in some crystal phases.
For this reason, and because of the lack of evidence that piezoelectricity is an important loss mechanism, we do not exclude any materials based on their piezoelectricity from our search.
However, we do provide the space group and whether or not it is centrosymmetric in Table~\ref{tab:results}, as it may be a consideration when designing a device.

From our survey, we can suggest materials that may be fruitful to explore.
Materials with a large Debye frequency will be best at suppressing the loss tangent.
We would also recommend choosing materials with a large band gap to potentially avoid the type of loss mechanisms that were observed in Si.
Diamond is by far the best choice: it has the largest Debye frequency and lowest loss tangent as a result.
Furthermore, it has a large band gap, avoiding the type of losses observed in Si, and is centrosymmetric, avoiding piezoelectric contributions.
Cubic BN is second best, its main drawback being piezoelectricity, which as we noted can be engineered around.
However, these materials may be difficult to work with and incorporate in devices.
An alternative to cubic BN may be found by considering the group-III nitrides (including InN, GaN, and AlN), a mature materials platform due to their use in solid-state lighting~\cite{zhou_reviewcurrent_2017}.
AlN has a band gap comparable to cubic BN and lower loss than Al$_2$O$_3$ (assuming identical defect chemistry).
There have been initial efforts to produce all-nitride qubits~\cite{nakamura_superconducting_2011,kim_enhanced_2021}.
Similarly, SiC has one of the lowest loss values within our model, yet is easier to work with for devices.

\section{Conclusions}

In conclusion, we have studied the impact of charged defects on dielectric loss in a variety of materials.
Charged defects break translational symmetry and enable the absorption of electromagnetic radiation by acoustic phonons, and we derived the absorption coefficient and loss tangent for this process.
Our calculated loss tangent of $7.2 \times 10^{-9}$ is in good agreement with measurements on high-grade Al$_2$O$_3$, which found a value of $19 \times 10^{-9}$~\cite{read_precision_2023}.
Furthermore, we expect the loss tangent to be independent of temperature for this process, in agreement with the observations.
This demonstrates that charged defects are likely limiting the bulk loss in such samples and controlling their incorporation is necessary to improve coherence times.
In addition, the loss is highly dependent on properties of the host material, in particular the Debye frequency.
We implemented a high-throughput search on a database of first-principles calculations with the goal of identifying alternative low-loss materials.
Based on our results, we suggest that diamond, cubic BN, AlN, and SiC are promising materials to explore.

\begin{acknowledgements}
    We gratefully acknowledge fruitful discussions with C. E. Dreyer, N. P. de Leon, and R. J. Schoelkopf.
    This work was supported by the U.S. Department of Energy, Office of Science, National Quantum Information Science Research Centers, Co-design Center for Quantum Advantage (C2QA) under contract number DE-SC0012704.
    The research used resources of the National Energy Research Scientific Computing Center, a DOE Office of Science User Facility supported by the Office of Science of the U.S. Department of Energy under Contract No. DE-AC02-05CH11231 using NERSC award BES-ERCAP0021021.
\end{acknowledgements}

\section*{Data Availability}
The data that support the findings of this study are openly available in Zenodo at \url{https://doi.org/[DOI-TO-BE-INSERTED]}.

\bibliographystyle{apsrev4-2}

\end{document}